\documentclass[aps,pra,reprint,longbibliography,superscriptaddress]{revtex4-2}

\usepackage[T1]{fontenc}
\usepackage[utf8]{inputenc}
\usepackage{amsmath,amssymb,amsfonts,bm,mathtools}
\usepackage{physics}
\usepackage{graphicx}
\usepackage{xcolor}
\usepackage{booktabs}
\usepackage{multirow}
\usepackage{hyperref}
\usepackage{enumitem}
\usepackage{comment}
\usepackage{ragged2e}
\usepackage{cleveref}
\usepackage{placeins}

\usepackage{orcidlink}

\usepackage{tikz}

\hypersetup{colorlinks=true,linkcolor=blue,citecolor=blue,urlcolor=blue}

\newcommand{\Ree}{\operatorname{Re}}
\newcommand{\Imm}{\operatorname{Im}}
\newcommand{\ii}{\mathrm{i}}
\newcommand{\coh}{\mathrm{coh}}
\newcommand{\TPM}{\mathrm{TPM}}
\newcommand{\MH}{\mathrm{MH}}
\newcommand{\KD}{\mathrm{KD}}
\newcommand{\STA}{\mathrm{STA}}
\newcommand{\CD}{\mathrm{CD}}
\newcommand{\ad}{\mathrm{ad}}
\newcommand{\off}{\mathrm{off}}

\begin{document}

\title{Shortcut-error signatures in coherence-retaining endpoint work quasistatistics}

\author{Gabriella G. Damas~\orcidlink{0000-0003-3376-9281}}
\affiliation{Department of Physics, Zhejiang Normal University, Jinhua 321004, China}
\affiliation{Instituto de Física, Universidade Federal de Goiás, 74.001-970, Goiânia
- GO, Brazil}

\author{G. D. de Moraes Neto~\orcidlink{0000-0003-4273-8380}}
 \email{gdmneto@gmail.com}
\affiliation{Faculty of Civil Engineering and Mechanics, Kunming University of Science and Technology, Kunming, 650500, China}

\date{\today}

\begin{abstract}
Quantum work statistics differ from classical ones because initial energy coherence matters. The standard two‑point measurement (TPM) gives a positive distribution but erases the initial phase information. Coherence‑retaining endpoint‑work quasistatistics as a compact probe of shortcut‑to‑adiabaticity performance. When work is defined with respect to a reference Hamiltonian at the start and end, an exact counterdiabatic shortcut pulls the final reference Hamiltonian back to an operator diagonal in the initial energy basis. Endpoint Kirkwood‑Dirac (or Margenau‑Hill) quasistatistics then lose sensitivity to initial coherence and reduce to the TPM result. Imperfect shortcuts restore this coherence sensitivity: a non‑commuting control error yields off‑diagonal pulled‑back Hamiltonian elements at first order in the error amplitude, whereas population‑only transition probabilities change only at second order. Harmonic‑oscillator and qubit benchmarks confirm this linear‑vs‑quadratic contrast. The result is complementary to inclusive work‑cost analyses of counterdiabatic driving: it does not measure the auxiliary field’s energetic cost, but provides a phase‑sensitive endpoint diagnostic of residual nonadiabaticity.
\end{abstract}

\maketitle

\section{Introduction}

Fluctuating work is central to nonequilibrium thermodynamics, but its quantum definition is constrained by measurement backaction. The standard two-point measurement (TPM) protocol measures the energy before and after a drive, producing a positive work distribution that satisfies classical-like fluctuation theorems \cite{Talkner2007,Esposito2009,Campisi2011,Batalhao2014,Talkner2016,Solfanelli2021}.
This operational clarity comes at a cost: the first projective measurement replaces the actual initial state by its dephased counterpart, thus removing coherences between distinct energy eigenspaces before the dynamics begins \cite{Perarnau2017,Lostaglio2018,Xu2018,Micadei2020,Landi2024}.

This erasure is not a technical artifact. No-go results show that one cannot construct a universal positive work distribution that both reproduces TPM for incoherent states and yields the correct average energy change for arbitrary coherent states \cite{Perarnau2017,Hovhannisyan2024}. Coherence-sensitive formulations therefore use characteristic functions or quasiprobabilities, including full-counting-statistics, Kirkwood--Dirac (KD), and Margenau--Hill (MH) constructions \cite{Allahverdyan2014,Solinas2015,Hofer2017,Lostaglio2018,Francica2022,Diaz2020,Pei2023,Lostaglio2023,Gherardini2024}.
Endpoint-measurement approaches were introduced to capture the influence of initial coherence on energy fluctuations without imposing the destructive initial projection \cite{Gherardini2021Endpoint}. Recent experiments have reconstructed KD- and MH-type work quasiprobabilities using interferometric and projective techniques \cite{HernandezGomez2024,HernandezGomez2024PRR}. These developments motivate a more specific question: how do coherence-retaining endpoint statistics respond to physically relevant control protocols?

Shortcuts to adiabaticity (STA) provide a natural setting for this question. Through counterdiabatic or transitionless driving, STA implements the eigenstate transport of a slow adiabatic reference process in finite time \cite{Berry2009,Torrontegui2013,GueryOdelin2019}. Because these protocols are used in quantum engines, state preparation, and information processing, their thermodynamic cost has been widely studied. Most analyses focus on inclusive work, where the auxiliary counterdiabatic Hamiltonian is included as part of the driven system; in that setting, mean auxiliary work can vanish while work fluctuations broaden and encode geometric information \cite{Zheng2016,Funo2017,Zhang2018}. The present Letter considers a different object: endpoint work defined only with respect to the reference Hamiltonians $H_0(0)$ and $H_0(\tau)$. We do not quantify the energetic cost of implementing $H_{\rm CD}(t)$. Instead, we use endpoint statistics to test whether the implemented unitary transports the final reference energy measurement into a form compatible with the initial one.

The central result connects the compatibility structure of KD/MH endpoint statistics to STA calibration. Exact transitionless driving enforces compatibility for the reference Hamiltonian, causing the coherence contribution to the endpoint work quasimean to vanish. A small non-commuting control error breaks this compatibility and generates an off-diagonal endpoint sector, producing a coherent endpoint signal that is linear in the error amplitude whenever the prepared coherence overlaps with that sector. By contrast, transition probabilities and population-only excesses scale quadratically. Thus initial quantum coherence becomes a first-order endpoint witness for non-commuting shortcut errors.

The diagnostic has three parts. First, the first moment, characteristic function, and fine-grained KD/MH weights are separated, clarifying the precise meaning of the collapse to the TPM-compatible result. Second, perturbation theory shows that an imperfect unitary $U_\epsilon=U_{\STA}e^{-\ii\epsilon K}+O(\epsilon^2)$ generates off-diagonal endpoint terms proportional to $[K,H_{\ad}]$. Third, benchmarks in a parametric oscillator and a driven qubit demonstrate the linear endpoint response and the revival of KD/MH nonclassical signatures, including imaginary KD weights and MH negativity.

\section{Endpoint coherence sector}
\label{sec:criterion}

\begin{figure}[t]
\centering
\includegraphics[width=0.98\linewidth]{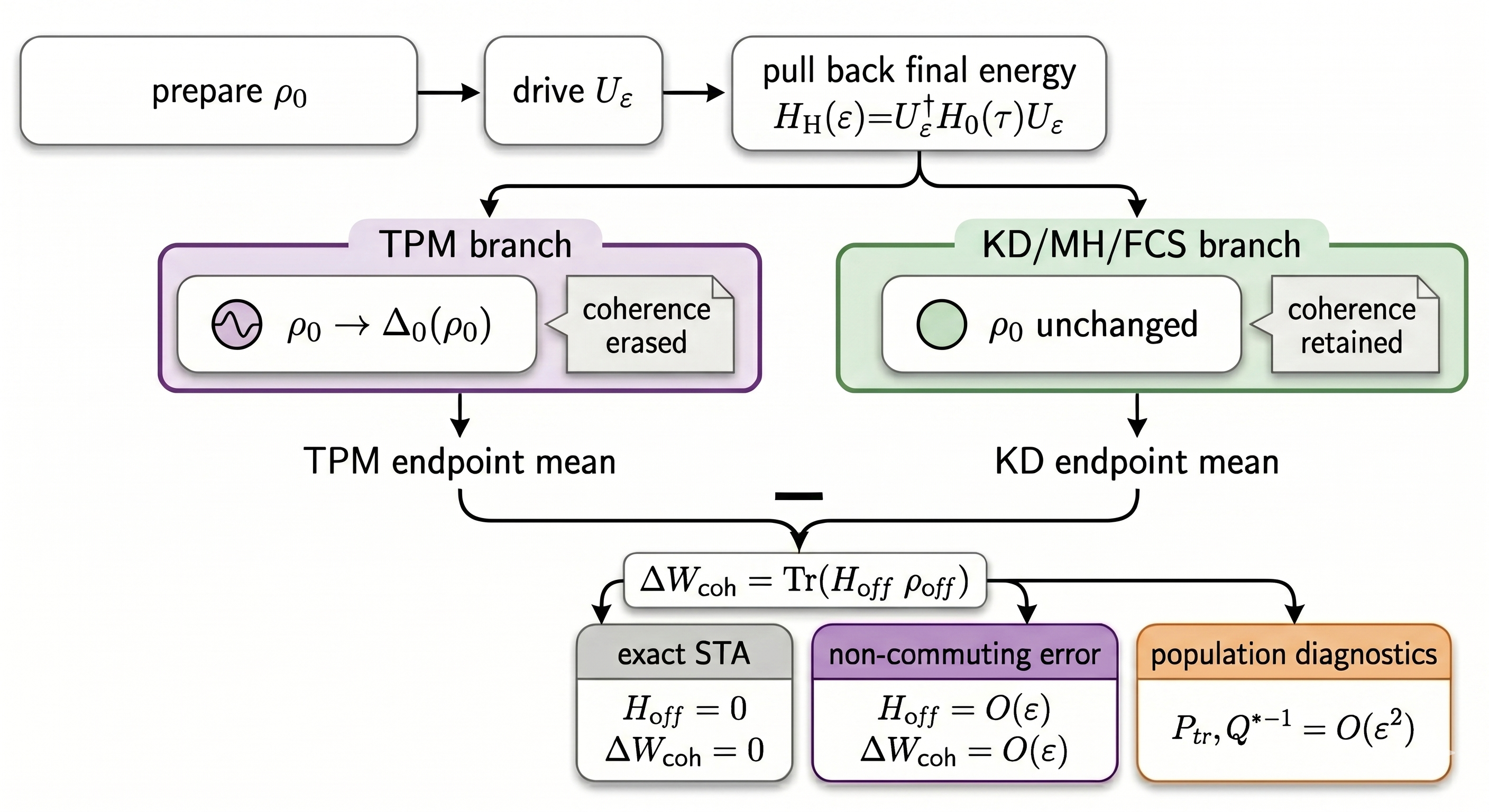}
\caption{Endpoint-work diagnostic flowchart. The traditional TPM branch applies an initial dephasing map $\Delta_0$, erasing energy-basis coherences. The Kirkwood-Dirac (KD) and Margenau-Hill (MH) branch retains the full initial state $\rho_0$. The difference between their endpoint means, $\Delta W_{\coh}$, isolates the nonadiabatic coherent response driven by the off-diagonal pulled-back Hamiltonian $H_{\off}$.}
\label{fig:compatibility}
\end{figure}

To distinguish the traditional thermodynamic approach from the quasiprobabilistic one, we establish the coherence-retaining endpoint framework. Let
\begin{equation}
H_0(0)=\sum_\alpha E_\alpha^0\Pi_\alpha^0,
\qquad
\Delta_0(\rho)=\sum_\alpha \Pi_\alpha^0\rho\Pi_\alpha^0,
\label{eq:dephasing}
\end{equation}
where $\Pi_\alpha^0$ are spectral projectors of the initial reference Hamiltonian. This form includes degeneracies; for a nondegenerate spectrum each projector is one-dimensional. For a closed implemented protocol $U$, define the final reference Hamiltonian pulled back to the initial frame,
\begin{equation}
H_H(\tau)=U^\dagger H_0(\tau)U .
\label{eq:HHdef}
\end{equation}
The TPM endpoint mean is
\begin{equation}
\langle W_0\rangle_{\TPM}=\Tr\{[H_H(\tau)-H_0(0)]\Delta_0(\rho_0)\},
\end{equation}
whereas the coherence-retaining endpoint quasimean is
\begin{equation}
\langle W_0\rangle_{\coh}=\Tr\{[H_H(\tau)-H_0(0)]\rho_0\}.
\end{equation}
The conceptual distinction between these two operational branches is summarized in Fig. \ref{fig:compatibility}. The second quantity is the first moment of a KD/MH/FCS endpoint quasistatistic \cite{Allahverdyan2014,Solinas2015}. It is not a positive joint-measurement average of two generally incompatible observables, and it does not evade no-go theorems \cite{Francica2022}. Equivalently, it is the expectation-value energy change obtained without applying the destructive initial energy measurement. The difference from TPM is
\begin{align} \nonumber
\Delta W_{\coh}
&=\langle W_0\rangle_{\coh}-\langle W_0\rangle_{\TPM} \\
&=\Tr\{H_H(\tau)[\rho_0-\Delta_0(\rho_0)]\}.
\label{eq:general_delta}
\end{align}
Thus only the off-diagonal sector of $H_H$ between initial energy blocks matters. Defining
\begin{equation}
H_{\off}=H_H-\Delta_0(H_H),
\qquad
\rho_{\off}=\rho_0-\Delta_0(\rho_0),
\end{equation}
we have
\begin{align} \nonumber
\Delta W_{\coh}&=\Tr(H_{\off}\rho_{\off}),
\\
|\Delta W_{\coh}|&\leq \|H_{\off}\|_\infty\,\|\rho_{\off}\|_1 .
\label{eq:norm_bound}
\end{align}
For a nondegenerate initial spectrum this gives the elementary estimate
\begin{equation}
|\Delta W_{\coh}|
\leq \max_{m\neq n}|(H_H)_{mn}|
\sum_{m\neq n}|(\rho_0)_{mn}|.
\label{eq:l1bound}
\end{equation}
The bound shows that both a coherent input state and a pulled-back final Hamiltonian with off-diagonal components are required.

The endpoint characteristic functions are
\begin{align}
\chi_\rho(u)&=\Tr\!\left[e^{\ii uH_H(\tau)}e^{-\ii uH_0(0)}\rho_0\right],\notag\\
\chi_{\Delta\rho}(u)&=\Tr\!\left[e^{\ii uH_H(\tau)}e^{-\ii uH_0(0)}\Delta_0(\rho_0)\right].
\label{eq:chis}
\end{align}
If $[H_H(\tau),\Pi_\alpha^0]=0$ for all $\alpha$, then $e^{\ii uH_H(\tau)}e^{-\ii uH_0(0)}$ is block diagonal in the same projectors as $\Delta_0$. Hence
\begin{equation}
\chi_\rho(u)=\chi_{\Delta\rho}(u)
\qquad \text{for all }u\text{ and all }\rho_0 .
\label{eq:chi_collapse}
\end{equation}
This proves the characteristic-function collapse, not merely the first-moment equality.

For the full fine-grained KD weights one must track projectors. Let $\Pi_m^\tau$ be final energy projectors and $\Pi_m^H=U^\dagger\Pi_m^\tau U$. Then
\begin{equation}
q_{m\alpha}^{\KD}=\Tr(\Pi_m^H\Pi_\alpha^0\rho_0),
\qquad
q_{m\alpha}^{\MH}=\Ree q_{m\alpha}^{\KD}.
\label{eq:kd_mh}
\end{equation}
The full KD/MH weights are independent of initial coherences if the pulled-back final projectors commute with the initial energy projectors, $[\Pi_m^H,\Pi_\alpha^0]=0$ \cite{Pei2023,Beyer2022}. This condition is automatic for exact transitionless dynamics in a nondegenerate reference spectrum and holds blockwise for degenerate eigenspaces. For degenerate spectra, all statements should be understood blockwise with respect to the spectral projectors of $H_0(0)$. Rotations inside a degenerate energy block do not affect endpoint energy statistics, provided the final reference Hamiltonian is proportional to the identity within the corresponding final block. Resolving holonomic structure inside degenerate manifolds would require additional observables and is outside the present endpoint-work analysis. Thus the present endpoint criterion diagnoses incompatibility between energy
blocks; holonomies acting entirely within a degenerate block are invisible to
energy-only endpoint statistics unless additional observables resolve the
internal block structure.

As summarized in Fig.~\ref{fig:compatibility}, the exact STA yields the compatible case. For transitionless driving in a
nondegenerate reference spectrum \cite{Torrontegui2013},
\begin{equation}
U_{\STA}|n(0)\rangle=e^{\ii\phi_n}|n(\tau)\rangle,
\end{equation}
so that
\begin{equation}
U_{\STA}^\dagger H_0(\tau)U_{\STA}
=\sum_nE_n^\tau |n(0)\rangle\langle n(0)|.
\label{eq:sta_diagonal}
\end{equation}
Thus the pulled-back final reference Hamiltonian is diagonal in the initial energy basis. Consequently, exact STA removes the coherence contribution to the endpoint-work quasimean and collapses the endpoint characteristic function to its dephased, TPM-compatible form. For nondegenerate spectra, the pulled-back final projectors are also compatible with the initial energy projectors, so the fine-grained KD/MH weights lose their dependence on initial energy coherence. 

When the shortcut is not exact, a small non-commuting shortcut error makes the pulled-back Hamiltonian acquire an off-diagonal correction at first order in the error amplitude. The benchmarks below show how this revived $H_{\off}$ produces a linear coherent endpoint signal, while population-only errors remain quadratic.

\section{Non-commuting shortcut errors}
\label{sec:perturbative}

Let the implemented unitary be written in the initial adiabatic frame as
\begin{equation}
U_\epsilon=U_{\STA}e^{-\ii\epsilon K}+O(\epsilon^2),
\label{eq:Ueps}
\end{equation}
where $K=K^\dagger$ is the error generator and $\epsilon\ll1$ is a dimensionless error amplitude. For a waveform error $\delta H(t)$ around the exact STA Hamiltonian, $K$ is the time integral of the error Hamiltonian in the interaction picture of the ideal STA evolution, up to the usual time-ordering corrections. Let
\begin{equation}
H_{\ad}=U_{\STA}^\dagger H_0(\tau)U_{\STA}
\end{equation}
be the diagonal endpoint operator associated with the exact shortcut. Then
\begin{align}
H_H^{(\epsilon)}
&=U_\epsilon^\dagger H_0(\tau)U_\epsilon\notag\\
&=e^{\ii\epsilon K}H_{\ad}e^{-\ii\epsilon K}+O(\epsilon^2) \\ \nonumber
&=H_{\ad}+\ii\epsilon[K,H_{\ad}]+O(\epsilon^2).
\label{eq:comm_expansion}
\end{align}
For nondegenerate levels,
\begin{equation}
\langle m|H_H^{(\epsilon)}|n\rangle
=\ii\epsilon(E_n^\tau-E_m^\tau)K_{mn}+O(\epsilon^2),
\qquad m\neq n.
\label{eq:offdiag_linear}
\end{equation}
For degenerate spectra the same statement holds blockwise: only matrix elements of $K$ connecting different endpoint energy blocks contribute to $H_{\off}$ at first order, while rotations inside a degenerate block commute with the block energy.
A non-commuting error, $[K,H_{\ad}]\neq0$, therefore produces an off-diagonal endpoint sector at first order. A coherence-sensitive endpoint signal appears whenever the prepared $\rho_{\off}$ overlaps with this sector \cite{Whitty2022}. Population-transfer diagnostics remain quadratic because transition amplitudes between distinct adiabatic labels are first order:
\begin{equation}
P_{n\to m}=\epsilon^2|K_{mn}|^2+O(\epsilon^3),
\qquad m\neq n.
\label{eq: prob}
\end{equation}
This is the origin of the linear-versus-quadratic contrast.

The diagnostic is selective, not universal. \Cref{tab:taxonomy} summarizes the relevant error classes. A commuting phase error does not generate a leading off-diagonal endpoint signal. An endpoint energy-scale error can shift energies at first order, but it is diagonal and is not a residual transition error. Dephasing suppresses the prepared coherences, and leakage outside the modeled Hilbert space requires an enlarged description.

\begin{table*}[t]
\centering
\caption{Selectivity of the endpoint-coherence diagnostic. The signal targets coherent, non-commuting shortcut errors; it is not a universal STA-quality metric.}
\label{tab:taxonomy}
\begin{tabular*}{\textwidth}{@{\extracolsep{\fill}}llll@{}}
\toprule
Error class & Endpoint off-diagonal response & Population response & Interpretation \\
\midrule
Missing CD amplitude & $O(\epsilon)$ if $[K,H_{\ad}]\neq0$ & $O(\epsilon^2)$ & detected \\
Transverse spurious control & $O(\epsilon)$ & $O(\epsilon^2)$ & detected \\
Commuting phase/calibration error & absent at leading order & often absent & invisible to this witness \\
Endpoint energy-scale error & diagonal $O(\epsilon)$ shift & not a transition & different calibration problem \\
Energy-basis dephasing & suppresses $\rho_{\off}$ & not equivalent & reduces signal \\
Leakage outside model space & not captured & may dominate & future work \\
\bottomrule
\end{tabular*}
\end{table*}

\section{Harmonic oscillator benchmark}
\label{sec:qho}

\begin{figure}[t]
\centering
\includegraphics[width=1\linewidth]{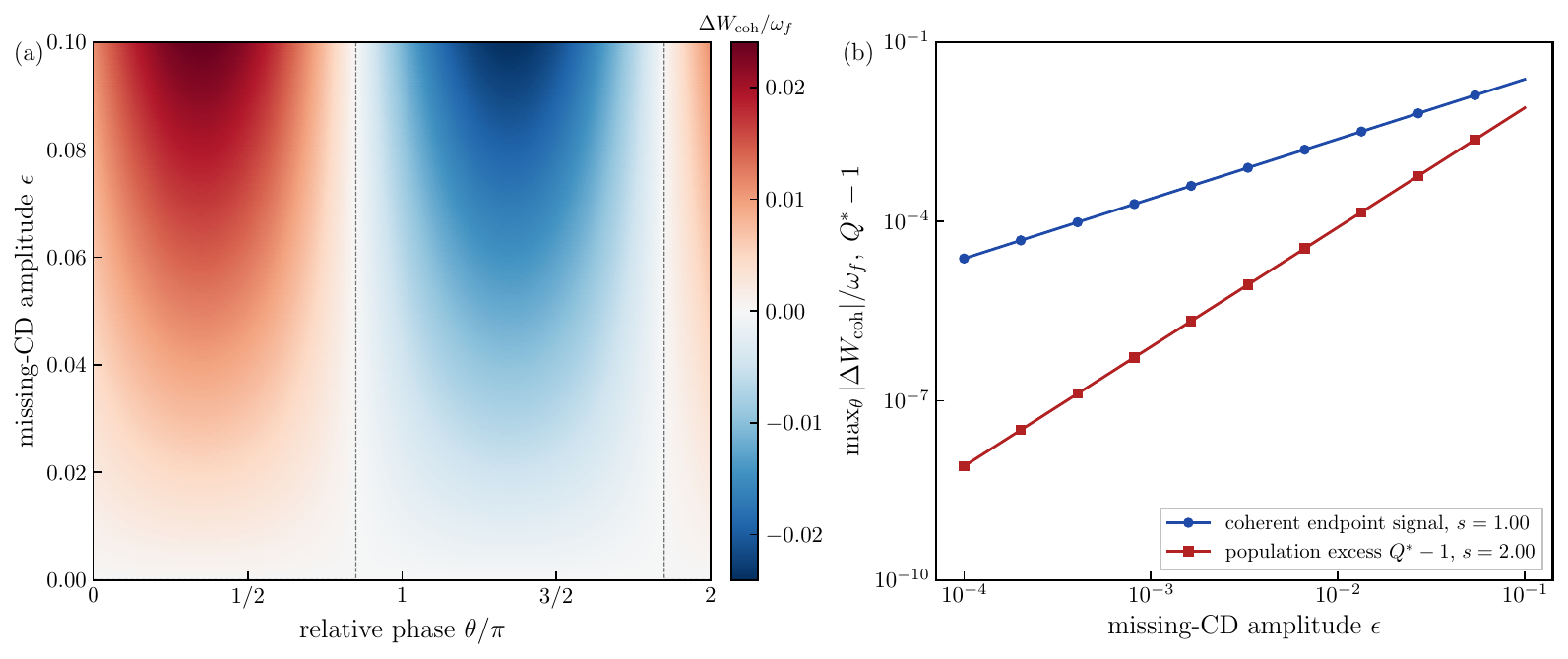}
\caption{Oscillator benchmark. (a) Coherent endpoint-work correction $\Delta W_{\mathrm{coh}}/\omega_f$ versus relative phase $\theta$ and missing-CD amplitude $\epsilon$ for the probe $|\psi_\theta\rangle=(|0\rangle+e^{i\theta}|2\rangle)/\sqrt{2}$. The TPM coherence contribution is identically zero. (b) Phase-optimized coherent signal $\max_\theta |\Delta W_{\mathrm{coh}}|$ extracted from (a) versus the population excess $Q^*-1$. Fits confirm the respective linear ($s=1.00$) and quadratic ($s=2.00$) error scalings.}
\label{fig:qho_phase}
\end{figure}

To demonstrate the selectivity and the linear scaling of the coherence-retaining diagnostic in a continuous-variable system, we first examine the time-dependent harmonic oscillator. This canonical model serves as a useful testbed because its linear dynamics allow for an exact analytical treatment of quadratic driving protocols via Bogoliubov transformations. The effect of a shortcut imperfection can be mapped directly onto a residual squeezing operator, providing a physical picture of the endpoint nonadiabaticity.

We set $m=\hbar=1$ and consider the reference Hamiltonian
\begin{equation}
H_0(t)=\frac{p^2}{2}+\frac{\omega^2(t)x^2}{2}.
\end{equation}
Let $a$ annihilate quanta of $H_0(0)$ and let $b$ annihilate quanta of $H_0(\tau)$. Any quadratic protocol generates a Bogoliubov transformation of the form
\begin{equation}
U^\dagger bU=\mu a+\nu a^\dagger,
\qquad |\mu|^2-|\nu|^2=1.
\label{eq:bog}
\end{equation}
The complex coefficients $\mu$ and $\nu$ encode the dynamics, with the off-diagonal amplitude $|\nu|^2$ quantifying the nonadiabatic excitations generated by the protocol. The pulled-back endpoint Hamiltonian is
\begin{equation}
H_H(\tau)=\omega_f\left[Q^*\left(a^\dagger a+\frac12\right)+Ca^{\dagger2}+C^*a^2\right],
\label{eq:HHqho}
\end{equation}
where
\begin{equation}
Q^*=|\mu|^2+|\nu|^2=1+2|\nu|^2,
\qquad C=\mu^*\nu .
\label{eq:QstarC}
\end{equation}
Therefore
\begin{equation}
\langle W_0\rangle_{\TPM}=\left(\omega_fQ^*-\omega_i\right)\left\langle a^\dagger a+\frac12\right\rangle,
\label{eq:TPMqho}
\end{equation}
and
\begin{equation}
\Delta W_{\coh}=\omega_f\left(C\langle a^{\dagger2}\rangle+C^*\langle a^2\rangle\right).
\label{eq:DeltaQHO}
\end{equation}
Thus $Q^*-1=2|\nu|^2$ is a second-order population excess, while $\Delta W_{\coh}$ is linear through $C=\mu^*\nu$.

For the phase probe
\begin{equation}
|\psi_\theta\rangle=\frac{|0\rangle+e^{\ii\theta}|2\rangle}{\sqrt2},
\qquad
\langle a^2\rangle=\frac{e^{\ii\theta}}{\sqrt2},
\end{equation}
we obtain
\begin{equation}
\Delta W_{\coh}=\sqrt2\omega_f\Ree(Ce^{-\ii\theta}).
\label{eq:phase_probe}
\end{equation}
The TPM baseline is independent of $\theta$. For coherent states,
\begin{equation}
\Delta W_{\coh}^{|\alpha\rangle}=2\omega_f\Ree(C^*\alpha^2),
\qquad
|\Delta W_{\coh}^{|\alpha\rangle}|\leq2\omega_f|\alpha|^2|C|.
\label{eq:coherent_bound}
\end{equation}

The oscillator counterdiabatic term is
\begin{equation}
H_{\CD}(t)=-\frac{\dot\omega}{4\omega}(xp+px),
\label{eq:qho_cd}
\end{equation}
which follows from the transitionless-driving construction for the parametric oscillator \cite{Berry2009,DelCampo2013,Torrontegui2013,GueryOdelin2019}. We use a quintic interpolation
\begin{align}
\omega(t)&=\omega_i+(\omega_f-\omega_i)s(t),
\\
s(t)&=10(t/\tau)^3-15(t/\tau)^4+6(t/\tau)^5,
\end{align}
with $\omega_i=1$, $\omega_f=2$, $\tau=\pi$, and $\dot\omega(0)=\dot\omega(\tau)=0$. The smooth endpoints avoid boundary kinks and make the residual response in the imperfect-STA case attributable to the missing counterdiabatic component. The implemented Hamiltonian is
\begin{equation}
H_\epsilon(t)=H_0(t)+(1-\epsilon)H_{\CD}(t).
\label{eq:qho_imperfect}
\end{equation}

The phase dependence in Eq. (\ref{eq:phase_probe}) raises the question of how to choose the probe phase $\theta$ without prior knowledge of the shortcut imperfection. If the dominant error mechanism is known (e.g., a systematic amplitude miscalibration), the phase of the residual squeezing axis $C$ is fixed, and $\theta$ can be set analytically to maximize the signal. For an uncharacterized error, the preparation phase can be swept empirically over $\theta \in [0, 2\pi]$. The amplitude of the resulting sinusoidal response yields the phase-optimized signal $\max_\theta |\Delta W_{\coh}|$, extracting the nonadiabatic error magnitude without \textit{a priori} alignment.

Figure \ref{fig:qho_phase} evaluates the phase dependence and error scaling for the oscillator. Figure \ref{fig:qho_phase}(a) shows the phase-sensitive coherence-retaining endpoint quasimean. The corresponding TPM coherence contribution is identically zero because the initial energy coherence is removed by the first projective measurement. Extracting the maximal coherent signal over $\theta$ gives the perturbative hierarchy shown in Fig. \ref{fig:qho_phase}(b): the coherent endpoint response scales linearly with the missing counterdiabatic amplitude $\epsilon$, whereas the nonadiabatic population excess $Q^*-1$ scales quadratically.

\begin{figure}[t]
\centering
\includegraphics[width=0.85\linewidth]{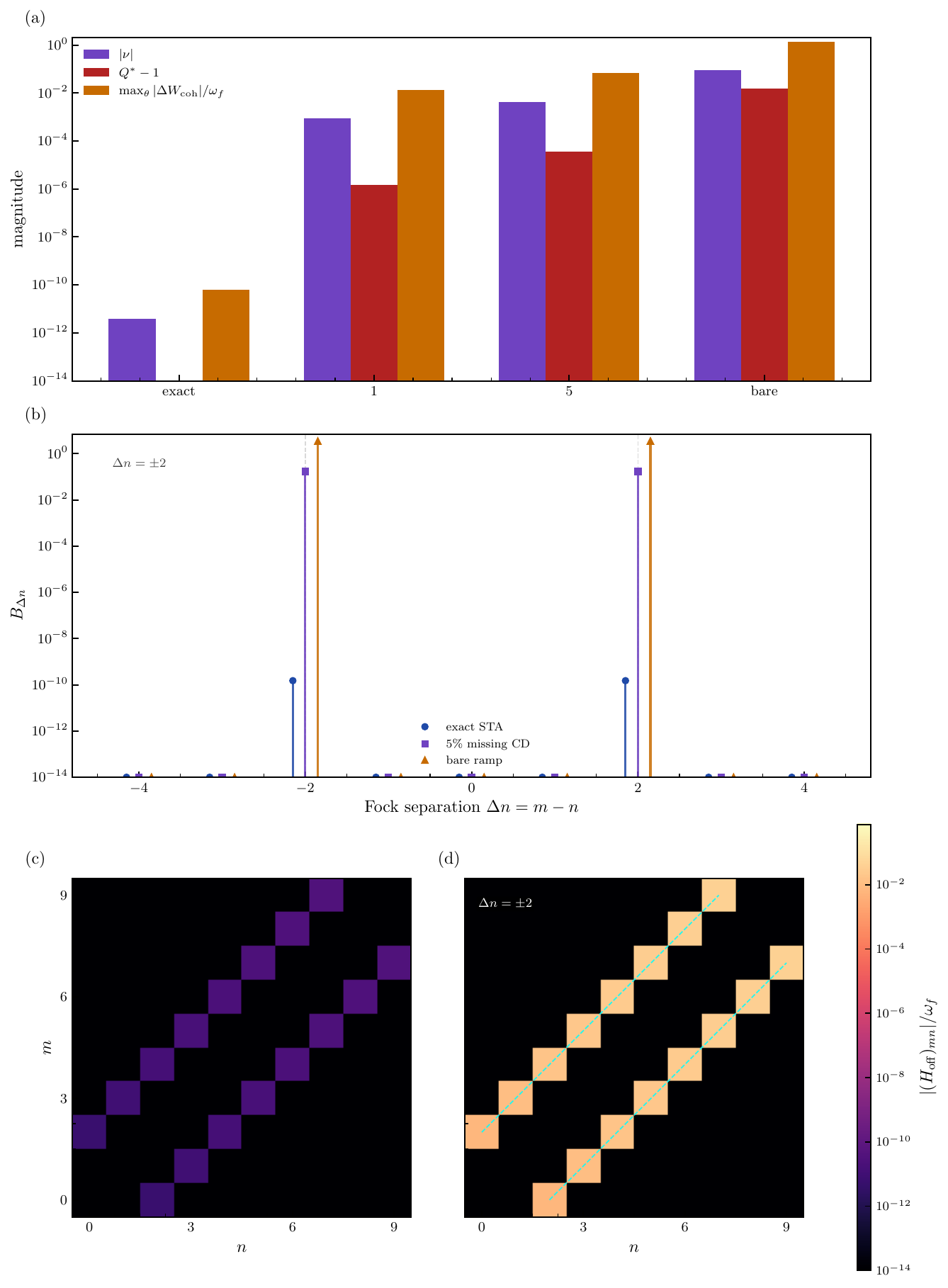}
\caption{Oscillator endpoint-error fingerprints. (a) Protocol-level diagnostics: residual Bogoliubov amplitude $|\nu|$, population excess $Q^*-1$, and phase-optimized coherent signal $\max_\theta|\Delta W_{\rm coh}|/\omega_f$ for exact STA, imperfect STA, and a bare ramp. (b) Band-resolved off-diagonal weight $B_{\Delta n}=\sum_n |(H_{\rm off})_{n+\Delta n,n}|/\omega_f$ isolating the $\Delta n=\pm2$ response. (c,d) Matrix representations of $|(H_{\rm off})_{mn}|/\omega_f$ in a truncated Fock basis ($n,m\leq9$) for exact STA and a $5\%$ missing-CD error.}
\label{fig:qho_operator}
\end{figure}

The physical origin of the linear signal is the off-diagonal sector of the pulled-back endpoint Hamiltonian. Figure \ref{fig:qho_operator}(a) compares the residual Bogoliubov amplitude $|\nu|$, the population excess $Q^*-1$, and the phase-optimized coherent endpoint signal for exact STA, imperfect STA, and a bare ramp. Exact transitionless driving suppresses all three quantities to the numerical precision floor. Imperfect shortcuts revive the off-diagonal Bogoliubov amplitude, and the coherent endpoint signal responds at first order.

The band-resolved profile in Fig. \ref{fig:qho_operator}(b) identifies the structure of this revival. Within this ideal quadratic oscillator benchmark, the physical off-diagonal sector has support on the squeezing bands $\Delta n=\pm2$, as expected from the $a^{\dagger2}$ and $a^2$ terms in Eq. (\ref{eq:HHqho}).
The full matrix fingerprints in Fig. \ref{fig:qho_operator}(c,d) provide the same information in the Fock basis: exact STA leaves only a numerical floor, whereas a $5\%$ missing-CD error revives the two squeezing-like off-diagonal bands. The apparent residual structure in the exact-STA panel is set by the plotting floor and numerical precision.

\FloatBarrier

\section{Qubit benchmark and quasiprobability revival}
\label{sec:qubit}

For the two-level benchmark we take
\begin{equation}
H_0(t)=\frac12\left[\Delta\sigma_x+\lambda(t)\sigma_z\right],
\end{equation}
with $\Delta=1$, $\lambda_i=-4$, $\lambda_f=4$, $\tau=6$, and the same quintic interpolation for $\lambda(t)$. The counterdiabatic term is \begin{equation}
H_{\CD}(t)=\frac12
\frac{\bm B(t)\times\dot{\bm B}(t)}{|\bm B(t)|^2}\cdot\bm\sigma,
\qquad
\bm B(t)=(\Delta,0,\lambda(t)).
\end{equation}
We implement
\begin{equation}
H_\epsilon(t)=H_0(t)+(1-\epsilon)H_{\CD}(t).
\end{equation}
Writing the pulled-back endpoint Hamiltonian as
\begin{equation}
H_H^{(\epsilon)}=\frac12\bm h_H^{(\epsilon)}\cdot\bm\sigma,
\end{equation}
and decomposing $\bm h_H$ into components parallel and perpendicular to the initial energy axis, an initial Bloch vector $\bm r_\perp$ transverse to that axis gives
\begin{equation}
\Delta W_{\coh}
=
\frac12\bm h_{H,\perp}^{(\epsilon)}\cdot\bm r_\perp .
\label{eq:qubit_bloch}
\end{equation}
The maximal equatorial signal is therefore $|\bm h_{H,\perp}^{(\epsilon)}|/2$. For a non-commuting shortcut error this quantity is first order in the residual error amplitude, whereas the transition probability from the initial ground state to the final excited state is quadratic.

The KD weights test quasiprobabilistic information beyond the first endpoint moment:
\begin{equation}
q_{m n}^{\KD}(\epsilon,\phi)
=
\Tr\!\left(\Pi_m^H(\epsilon)\Pi_n^0\rho_\phi\right),
\qquad
\rho_\phi=|\psi_\phi\rangle\langle\psi_\phi|,
\label{eq:qubit_kd}
\end{equation}
where
\begin{equation}
|\psi_\phi\rangle
=
\frac{|g_0\rangle+e^{\ii\phi}|e_0\rangle}{\sqrt2}.
\end{equation}
The corresponding TPM weights are obtained from the same imperfect protocol after dephasing the initial state,
\begin{equation}
p_{mn}^{\TPM}(\epsilon,\phi)
=
\Tr\!\left(\Pi_m^H(\epsilon)\Pi_n^0\Delta_0(\rho_\phi)\right).
\end{equation}
We monitor the phase-optimized KD deviation
\begin{equation}
D_{\KD}^{\max}(\epsilon)
=
\max_\phi
\sum_{mn}
\left|
q_{mn}^{\KD}(\epsilon,\phi)
-
p_{mn}^{\TPM}(\epsilon,\phi)
\right|,
\end{equation}
and the phase-optimized Margenau--Hill negativity
\begin{equation}
\mathcal N_{\MH}^{\max}(\epsilon)
=
\max_\phi
\sum_{mn}
\max\!\left[0,-\Ree q_{mn}^{\KD}(\epsilon,\phi)\right].
\end{equation}
The quantity $D_{\KD}^{\max}$ measures the coherence-induced departure from the TPM weights, whereas $\mathcal N_{\MH}^{\max}$ detects negative real KD/MH weights. The imaginary KD matrix is evaluated at the phase maximizing $\sum_{mn}|\Imm q_{mn}^{\KD}|$.

\begin{figure}[t]
\centering
\includegraphics[width=0.98\linewidth]{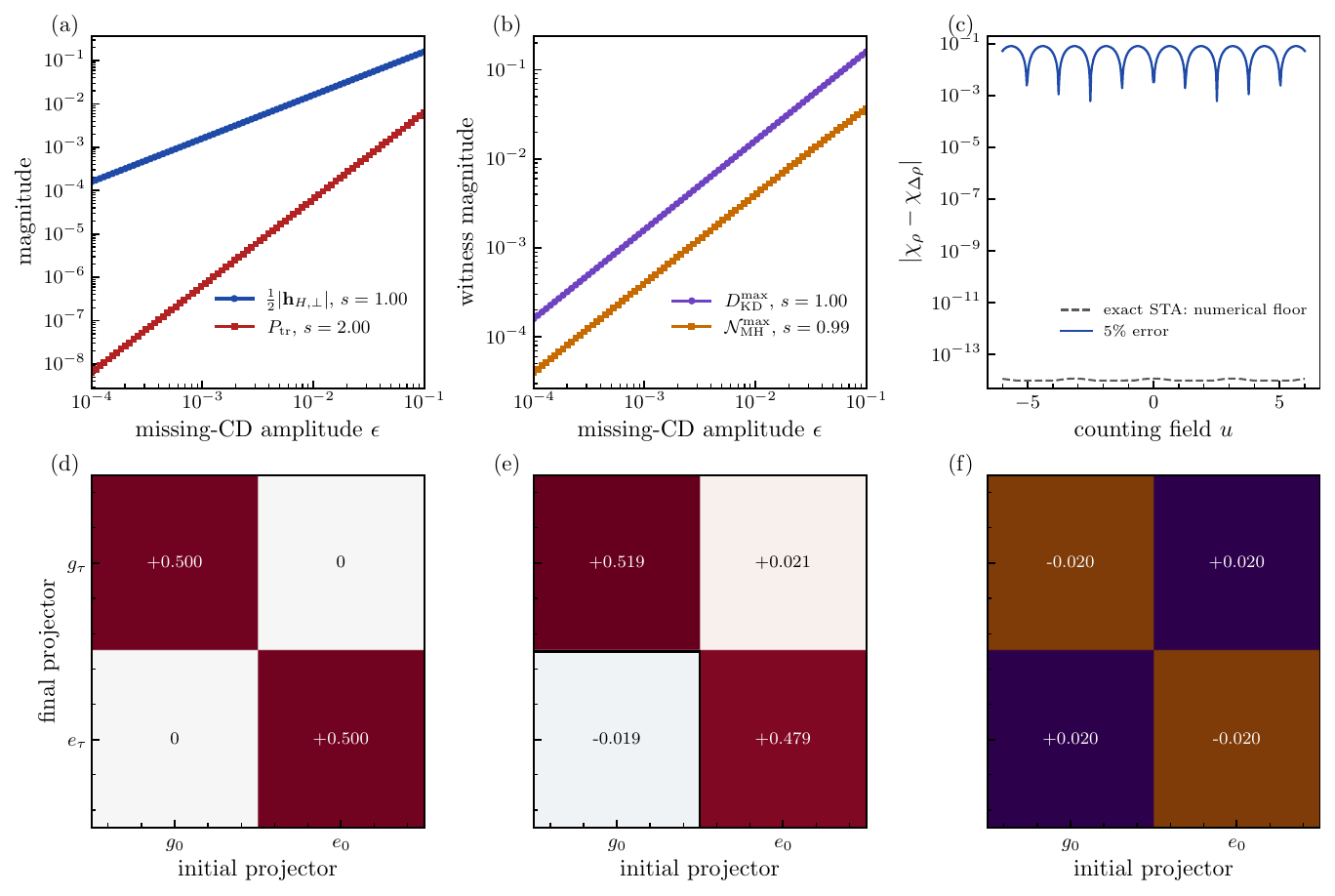}
\caption{
Qubit benchmark of endpoint quasistatistics.
(a) Transverse endpoint field $\frac{1}{2}|\mathbf h_{H,\perp}|$ and transition probability $P_{\rm tr}$ versus missing-CD amplitude $\epsilon$, showing linear and quadratic scaling, respectively.
(b) Phase-optimized quasiprobability witnesses $D_{\rm KD}^{\max}$ and $\mathcal N_{\rm MH}^{\max}$ versus $\epsilon$.
(c) Characteristic-function gap $|\chi_{\rho}(u)-\chi_{\Delta\rho}(u)|$ for exact STA and a $5\%$ missing-CD error; the exact-STA curve is limited by the numerical floor.
(d--f) KD/MH matrix fingerprints: (d) real KD weights for exact STA; (e) real KD/MH weights for a $5\%$ error at the phase maximizing $\mathcal N_{\rm MH}$; and (f) imaginary KD weights for a $5\%$ error at the phase maximizing $\sum_{mn}|{\rm Im}\,q^{\rm KD}_{mn}|$.
Matrix labels denote final and initial energy projectors.
}
\label{fig:qubit_kd}
\end{figure}

Figure \ref{fig:qubit_kd} summarizes the endpoint diagnostic for this qubit benchmark.  Figure \ref{fig:qubit_kd}(a) shows the maximum coherent endpoint signal,
\begin{equation}
\max_{|\bm r_\perp|=1}|\Delta W_{\coh}|
=
\frac12|\bm h_{H,\perp}^{(\epsilon)}|,
\end{equation}
alongside the TPM-style transition probability from the initial ground state to the final excited state. Fits over the weak-error window confirm that the coherent endpoint signal is linear in the residual error, while the population transition probability remains quadratic.

Figure \ref{fig:qubit_kd}(b) demonstrates that the quasiprobabilistic witnesses revive with the same perturbative hierarchy. Both $D_{\KD}^{\max}$ and $\mathcal N_{\MH}^{\max}$ scale linearly with the shortcut imperfection. Thus the first-order response is not restricted to the endpoint mean; it also appears in the KD/MH quasiprobability structure. The characteristic-function collapse predicted by Eq. \ref{eq:chi_collapse} is verified in Fig. \ref{fig:qubit_kd}(c): the gap $|\chi_\rho(u)-\chi_{\Delta\rho}(u)|$ vanishes for exact STA up to the numerical floor, but becomes finite under a $5\%$ missing-CD error.

Figure \ref{fig:qubit_kd}(d--f) displays the underlying quasiprobability weights. Exact STA produces positive diagonal KD weights Fig. \ref{fig:qubit_kd}(d), reflecting compatibility between the pulled-back final and initial energy projectors. At $\epsilon=5\%$, the real KD/MH matrix evaluated at the phase maximizing $\mathcal N_{\MH}$ develops a negative entry Fig. \ref{fig:qubit_kd}(e). The imaginary KD matrix, evaluated at the phase maximizing $\sum_{mn}|\Imm q_{mn}^{\KD}|$, displays nonzero complex quasiprobability components Fig. \ref{fig:qubit_kd}(f). These matrix fingerprints connect the residual nonadiabaticity to the emergence of nonclassical endpoint-work quasistatistics.

\FloatBarrier

\section{Operational access and comparison with other diagnostics}
\label{sec:experiment}

For the first endpoint moment, a full quasiprobability reconstruction is not required. In a qubit experiment one may prepare an equatorial state in the initial energy basis, apply the implemented shortcut, and measure the final reference energy. Repeating the experiment with the dephased preparation $\Delta_0(\rho_0)$ gives the TPM-compatible branch, and the difference between the two endpoint energy means estimates $\Delta W_{\coh}(\phi)$. The dephased preparation can be realized by classical mixing over the initial energy populations, phase randomization, or an explicit dephasing operation in the initial energy basis.

This targeted first-moment protocol is much less demanding than reconstructing the full error map. Full process tomography scales as a $d^4$-parameter characterization problem, while full state tomography scales as a $d^2$-parameter characterization problem. Reconstructing the full fine-grained KD quasidistribution requires an overhead that grows with the number of initial--final energy pairs and with the measurement scheme used to access their complex correlations. By contrast, the endpoint coherent correction requires estimating one observable, the final reference Hamiltonian $H_0(\tau)$, for a chosen coherent preparation and its dephased counterpart. Thus the protocol avoids reconstructing a high-dimensional state, process, or full quasidistribution. The remaining cost is the platform-dependent cost of preparing the two input ensembles and measuring the endpoint energy expectation with sufficient precision. For oscillator platforms, the endpoint energy can be estimated either from final quadrature moments,
$\langle H_0(\tau)\rangle=\langle p^2\rangle/2+\omega_f^2\langle x^2\rangle/2$, or from number-resolved populations, $\langle H_0(\tau)\rangle=\omega_f\sum_n(n+1/2)p_n$, depending on the available
readout.

The work-quasistatistical language remains useful because it embeds this first moment into a broader hierarchy that contains characteristic functions, higher moments, imaginary KD components, and MH negativity. If the full endpoint quasidistribution is desired, one can access characteristic functions using ancilla-assisted Ramsey interferometry \cite{Dorner2013,Mazzola2013,Solinas2015}. Related work-quasiprobability and projective-measurement reconstructions have been implemented or demonstrated in NMR, NV-center, and superconducting-qubit platforms \cite{Batalhao2014,HernandezGomez2024,HernandezGomez2024PRR,Zhang2018}. The recent NV-center KD reconstruction is close in spirit because it accesses work quasiprobabilities and their first moments through interferometric correlation measurements \cite{HernandezGomez2024}.

A shot-noise estimate gives the scale of the first-moment protocol. If the final qubit energy outcomes are bounded by $\pm\Omega_f/2$ and the coherent signal is $|\Delta W_{\coh}|=A\epsilon$, then estimating the two endpoint energy means with $N_{\rm br}$ shots per branch gives the conservative variance bound
\begin{equation}
{\rm Var}(\widehat{\Delta W}_{\coh})
\leq
\frac{\Omega_f^2}{2N_{\rm br}} .
\end{equation}
Resolving the signal with signal-to-noise ratio $R$ therefore requires
\begin{equation}
N_{\rm br}
\gtrsim
\frac{R^2\Omega_f^2}{2A^2\epsilon^2},
\label{eq:shots}
\end{equation}
up to order-one factors depending on the allocation of shots and the actual outcome variances. Energy-basis dephasing during the protocol reduces the coherent amplitude as $A\to A e^{-\Gamma_\phi\tau}$, giving
\begin{equation}
N_{\rm br}
\gtrsim
\frac{R^2\Omega_f^2}{2A^2e^{-2\Gamma_\phi\tau}\epsilon^2}.
\end{equation}
This estimate is conservative and is not a platform-level noise budget. It identifies the expected advantage over population diagnostics near the high-fidelity regime, where coherent endpoint signals scale as $\epsilon$ while transition probabilities scale as $\epsilon^2$. The present construction can also be viewed as the two-time endpoint sector of broader temporal KD or process-quasiprobability frameworks; here we restrict to this boundary version because it is sufficient for diagnosing STA endpoint compatibility and avoids reconstructing a full multi-time process.

The diagnostic is not a universal error detector. A suitably chosen phase-sensitive observable, direct tomography of the pulled-back endpoint operator $H_H$, Loschmidt-echo probes, or fidelity-based diagnostics can also detect non-commuting errors. The present method is not claimed to be universally optimal. Its role is more specific: it measures the overlap between a chosen coherence sector and the off-diagonal endpoint Hamiltonian. Compared with tomography, it is less complete but requires only endpoint energy expectations; compared with fidelity-based probes, it is directly tied to endpoint work quasistatistics. Additional robustness checks for dephasing and for a smooth waveform-distortion error are given in Appendix~\ref{app:robustness}.

\section{Discussion and limitations}
\label{sec:discussion}

The result is a compatibility statement. Coherence-retaining endpoint statistics reduce to their TPM-compatible form when the initial energy measurement and the pulled-back final energy measurement commute. Exact STA enforces this compatibility for the reference Hamiltonian. Non-commuting shortcut imperfections break it, generate an off-diagonal endpoint sector, and can produce a coherent signal linear in the error amplitude. The aim is not to introduce a new general theory of work quasiprobabilities, but to use their known incompatibility structure as an STA-specific endpoint diagnostic.

The distinction from inclusive STA work is essential. Inclusive work accounts for the full implemented Hamiltonian, including the counterdiabatic control. Here we use only the reference Hamiltonians $H_0(0)$ and $H_0(\tau)$. The diagnostic therefore does not measure the energetic cost of the auxiliary field. Instead, it tests whether the implemented unitary transports the final reference energy measurement into a form compatible with the initial one. It is relevant to finite-time state transfer, quantum-engine strokes, adiabatic gates, and pulse validation, but it does not replace inclusive-work, control-cost, or speed-limit analyses.

The present treatment is closed and unitary. Open-system effects can suppress or mimic the signal. For simple energy-basis dephasing, $\rho_{mn}\to e^{-\Gamma_\phi\tau}\rho_{mn}$ for $m\neq n$, so $\Delta W_{\coh}$ is exponentially reduced. Leakage outside the modeled Hilbert space, strong degeneracies, non-Abelian holonomies, and detector backaction require separate platform-specific analysis.

Although the benchmarks are a harmonic oscillator and a qubit, the motivation is strongest in larger Hilbert spaces, where full tomography or full quasiprobability reconstruction becomes costly. The first endpoint moment only requires estimating the endpoint reference energy for a coherent preparation and for its dephased counterpart. This avoids reconstructing the full state, process, or quasidistribution, although the preparation, readout, and phase-scan costs remain platform dependent.

The Letter provides a bounded contribution: an endpoint compatibility criterion, a perturbative linear-versus-quadratic scaling law for non-commuting shortcut errors, and solvable benchmarks showing KD/MH quasiprobability revival. The witness is selective: it detects coherent non-commuting shortcut errors, but it does not replace population, leakage, control-cost, or full-noise diagnostics.

\section*{Data and code availability}
The numerical scripts used to generate the figures propagate the closed oscillator and qubit benchmarks and compute KD/MH quantities for the qubit. They are available upon request.

\section*{Acknowledgments}
G.G.D thanks Professor Gao Xialong for his kind hospitality and support during their stay at the Zhejiang Normal University. This work was supported by the Coordenação de Aperfeiçoamento de Pessoal de Nível Superior - Brasil (CAPES) - Finance Code 001.

\subsection*{Artificial Intelligence Usage Declaration}
In accordance with journal guidelines, the authors declare the use of Artificial Intelligence Generated Content (AIGC) tools during the preparation of this manuscript. Specifically, Gemini/ChatGPT was utilized for language polishing, text editing, and assisting with deep literature research. Additionally, Gemini was used to assist in the conceptualization and drafting of the schematic diagram (Figure 1). Following the use of these tools, the authors rigorously reviewed, modified, and validated all generated text, research insights, and visual content. The authors assume full and sole responsibility for the integrity, accuracy, and originality of the final manuscript and affirm that no AI tool fulfills the role of, nor is listed as, an author.

\begin{figure}[t]
\centering
\includegraphics[width=0.72\linewidth]{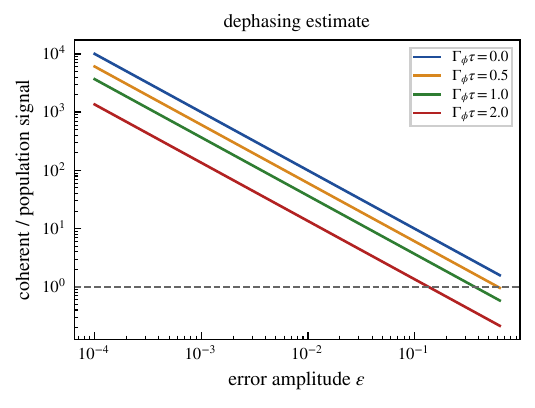}
\caption{ Simple dephasing estimate for the first-moment endpoint diagnostic. The plotted ratio compares a coherent signal proportional to $e^{-\Gamma_\phi\tau}\epsilon$ with a population signal proportional to
$\epsilon^2$. Dephasing reduces the coherent contrast but does not change the asymptotic linear-versus-quadratic hierarchy.}
\label{fig:S1_dephasing}
\end{figure}

\appendix
\section{Robustness checks}
\label{app:robustness}

This appendix collects two minimal robustness checks supporting the operational discussion in Sec.~VI. They are not intended as a platform-level noise model. Their role is to show that the linear-versus-quadratic hierarchy is not removed by simple coherence loss and is not specific to the missing-CD amplitude parametrization.

\begin{figure}[t]
\centering
\includegraphics[width=0.72\linewidth]{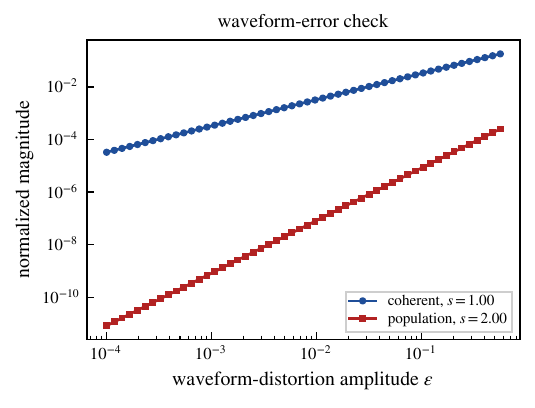}
\caption{Waveform-error check. For a smooth distortion of the shortcut waveform, the coherent endpoint signal scales linearly with the distortion amplitude, while the population-transfer diagnostic scales quadratically. This confirms that the linear-versus-quadratic hierarchy is a consequence of the non-commuting endpoint error sector, not of the special missing-CD parametrization.}
\label{fig:S2_waveform}
\end{figure}

First, consider pure dephasing in the initial energy basis. At the level of the first endpoint moment,
\begin{equation}
\rho_{mn}\rightarrow e^{-\Gamma_\phi\tau}\rho_{mn},
\qquad m\neq n,
\end{equation}
so that
\begin{equation}
|\Delta W_{\rm coh}|
\rightarrow
e^{-\Gamma_\phi\tau}|\Delta W_{\rm coh}|.
\end{equation}
If the closed-system coherent and population signals scale as
\begin{align} \nonumber
S_{\rm coh}&=A\epsilon,
\\
S_{\rm pop}&=B\epsilon^2,
\end{align}
then dephasing changes their ratio to
\begin{equation}
\frac{S_{\rm coh}}{S_{\rm pop}}
=
\frac{A}{B}\frac{e^{-\Gamma_\phi\tau}}{\epsilon}.
\end{equation}
This estimate is shown in Fig.~\ref{fig:S1_dephasing}. Dephasing reduces the coherent contrast, but the coherent signal remains parametrically larger than the population signal in the sufficiently small-error regime.

Second, we check that the scaling hierarchy is not an artifact of modeling the imperfection only as a missing-CD amplitude. A smooth waveform distortion also produces a first-order non-commuting correction to the pulled-back endpoint Hamiltonian whenever the induced error generator does not commute with $H_{\rm ad}$. The numerical check in Fig.~\ref{fig:S2_waveform} shows that the coherent endpoint signal remains linear in the waveform-distortion amplitude, whereas the population-transfer response remains quadratic.

\bibliographystyle{elsarticle-num}
\bibliography{references}

\end{document}